# Identification of Conduit Countries and Community Structures in the Withholding Tax Networks

**Tembo Nakamoto・Yuichi Ikeda**

**Abstract** Due to economic globalization, each country's economic law, including tax laws and tax treaties, has been forced to work as a single network. However, each jurisdiction (country or region) has not made its economic law under the assumption that its law functions as an element of one network, so it has brought unexpected results. We thought that the results are exactly international tax avoidance. To contribute to the solution of international tax avoidance, we tried to investigate which part of the network is vulnerable. Specifically, focusing on treaty shopping, which is one of international tax avoidance methods, we attempt to identified which jurisdiction are likely to be used for treaty shopping from tax liabilities and the relationship between jurisdictions which are likely to be used for treaty shopping and others. For that purpose, based on withholding tax rates imposed on dividends, interest, and royalties by jurisdictions, we produced weighted multiple directed graphs, computed the centralities and detected the communities. As a result, we clarified the jurisdictions that are likely to be used for treaty shopping and pointed out that there are community structures. The results of this study suggested that fewer jurisdictions need to introduce more regulations for prevention of treaty abuse worldwide.

## 1 Introduction

The financial crisis that occurred in 2008 not only brought about a once-in-a-100-years recession in terms of economic activity, it also represented a once-in-a-100-years turning point in international taxation. In dealing with the financial crisis, European countries were forced to raise consumption tax rates to enable large-scale fiscal spending, while the media and nonprofit organizations focused on bringing to light international tax avoidance by multinational companies (Inman 2016; Americans for Tax Fairness 2015). With economic globalization, international tax avoidance has increased with the facilitation of cross-border transactions.

Since tax treaties are basically concluded between two jurisdictions[1], it has hitherto been considered that problems relating to tax treaties should be discussed between the two contracting jurisdictions. However, it has been found that using a particular bilateral tax treaty in international tax avoidance erodes the tax sources of jurisdictions that did not conclude the bilateral tax treaty. As a result, vulnerabilities in one jurisdiction's tax treaty potentially erodes other jurisdictions' tax bases. For example, it is said that the reason why Mongolia did not conclude a tax treaty with the Netherlands is that Mongolia was worried that its tax base would erode under a tax treaty with the European country (McGauran 2013).

[1] The meaning of "jurisdiction" is almost the same as that of "country and region."

Economic globalization has significant implications for multinationals' business activities and international economic law. The activities of multinationals cross more borders and become more complex. In 1975, there were just 1,200 tax treaties worldwide; today, there are 3,200. In light of this, new thinking is beginning to emerge in the field of international tax law. It seeks to understand each jurisdiction's economic laws, including tax laws and tax treaties, as a single legal system (Ogata 2016).

Meanwhile, in recent years, network science has made remarkable progress by offering accurate network information in various fields. Although each field is different, finding networks follows the same principle (Barabási 2016). With this advance, network science has revealed interesting results that had hitherto not been found in a broad realm from sociology to physics. This study is an attempt to highlight an interesting aspect that has not yet been found far by making use of the knowledge of network science in the realm of international taxation, focusing particularly on the issue of international tax avoidance.

In the field of international taxation, as prior studies incorporating the method of network science, some studies have analyzed the withholding tax rate on dividends (Polak S. 2014; Van't Riet et al. 2014; Hong 2017), but they only studied withholding tax on dividends. The purpose of this study is to clarify "Treaty Shopping," which multinationals are said to use as one international tax avoidance scheme, from the viewpoint of tax rates and using the methods of network science. We study all important passive income, that is, dividends, interest, and royalties, and look for the relationships among each jurisdiction.

## 2 International Tax Avoidance

### 2.1 International Tax System

Companies are subject to various forms of taxation in the process of conducting cross-border transactions, most importantly corporate tax and withholding tax. According to international custom, each jurisdiction has the primary right to tax income which arises in or is derived from that jurisdiction (Arnold 2016). Therefore, each jurisdiction imposes corporate tax on business income arising in its jurisdiction and imposes withholding tax on dividends, interest, and royalties derived from its jurisdiction and paid to other jurisdictions. Even if dividends, interest and royalties are paid to companies in the same group, withholding tax is imposed as long as the group company is in another jurisdiction.

However, when dividends, interest, and royalties are paid to jurisdictions that conclude a tax treaty, the amount of withholding tax is often lower than usual. This is because many tax treaties have provisions to reduce the withholding tax rate or exempt the company from paying the tax altogether. This reduction or exemption benefit is generally called a tax treaty benefit. The purpose of offering a tax treaty benefit is to encourage economic activities between the two contracting jurisdictions, so the reduction or exemption is applied only to dividends, interest, and royalties paid to the other contracting state.

## 2.2 Treaty Shopping

The tax treaty benefits considered, the amount of withholding tax may be reduced by circumventing the third jurisdiction rather than paying dividends, interest, and royalties directly to other jurisdictions. For example, suppose that jurisdiction A (the source jurisdiction) imposes a 25% withholding tax, and jurisdiction C imposes a 5% withholding tax (Figure 1). Because jurisdiction A has concluded a tax treaty with jurisdiction C, the withholding tax is waived for dividends, interest and royalties paid from jurisdiction A to jurisdiction C. In this case, if a company located in jurisdiction A pays dividends, interest, and royalties directly to a company located in jurisdiction B, a 25% withholding tax are imposed. However, if a company located in jurisdiction A pays its dividends, interest, and royalties to a company located in jurisdiction B via jurisdiction C (the conduit jurisdictions), it is exempted from the withholding tax for payments made between jurisdiction A and jurisdiction C because of the A = C tax treaty, and jurisdiction C imposes a 5% withholding tax on payments made between jurisdiction C and jurisdiction B. In a case where a company located in jurisdiction A pays its dividends, interest, and royalties to a company located in jurisdiction B, if it pays them via jurisdiction C, it can avoid a 20% withholding tax, compared to paying them directly.

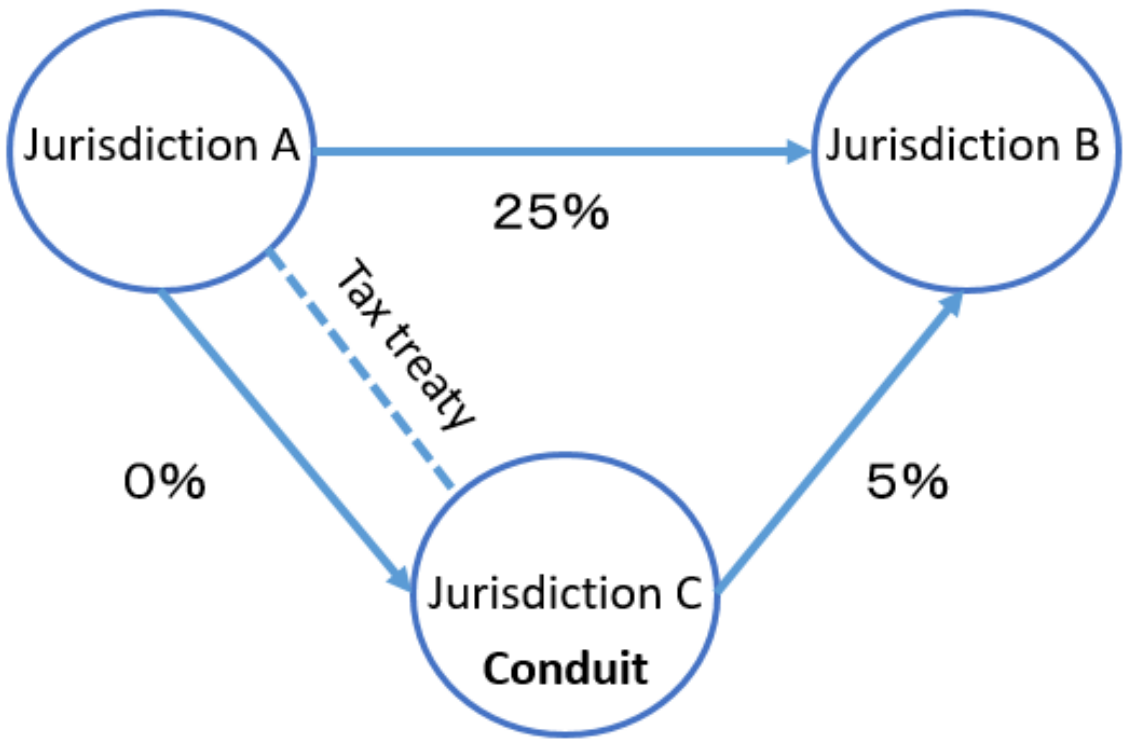

**Fig. 1** Treaty shopping

Reducing the amount of tax through the unexpected use of tax treaty benefits described above is called "Treaty Shopping." treaty shopping is not illegal unless violates the provisions of a jurisdiction's tax law or tax treaty, but particularly if it is done solely to reduce the amount of tax, it is called "treaty abuse" (Marian 2016) or "improper use of the convention" (OECD 2015) and should be corrected.

For companies engaged in treaty shopping, this is just one of the ways to avoid withholding tax. However, for source jurisdictions, this is one of the ways in which their rights to impose taxation are eroded by third jurisdictions (conduit jurisdictions). It is source jurisdictions that originally have the right to impose taxation since the dividends, interest, and royalties are income derived from their jurisdiction. Nevertheless, it is the third jurisdictions that impose their tax on this income. This is why treaty shopping is regarded as a form of tax base erosion and is viewed as a problem in the area of international taxation.

In this way, when the tax laws and tax treaties of multiple countries overlap, opportunities for an unexpected reduction in the tax burden may be created. This is called international tax avoidance. We see international tax avoidance as a problem emerging in an environment where economic globalization makes the economic legal system of each country one system as a whole. This is because we use the method of network science for international taxation.

## 3 Withholding Tax Network

### 3.1 Data

The data relating to the withholding tax rate of dividends, interest, and royalties comes from Ernst Young (2017) and Diamond (2017a). The 165 jurisdictions subject to this study are as shown in Table 1. With respect to the withholding tax rate on dividends, many tax treaties apply different tax rates depending on the percentage of holdings. Generally, the higher the shareholding ratio, the lower the tax rate. The interest in this study is in the tax burden that arises when the profit is transferred within a corporate group. This is because international tax avoidance by multinationals is often done through the transfer of their profits between group companies. Therefore, assuming that the dividend was generated by a 100% wholly owned subsidiary, the lowest tax rate are applied. Regarding interest, some jurisdictions have applied different tax rates for non-bank deposits and bank deposits. For the reasons mentioned above, it is assumed that the tax rate for the non-bank deposit is applied.

Some tax treaties have other requirements for granting withholding tax exemptions or reductions, apart from the shareholding ratio mentioned above. We consider that all requirements are met in this case. Moreover, the applied withholding tax rate is not always clear because the wording in a given tax treaty and the relationship between domestic tax laws and tax treaties are not always clear. In these cases, the lowest tax rate among the possible withholding tax rates is applied. On the other hand, although some jurisdictions do not impose withholding tax according to their domestic tax laws, their tax treaties determine the withholding tax rate. Since the primary purpose of tax treaties is usually to avoid international double taxation, it is considered that withholding tax is not imposed, but only in this case, we assume that the tax rate stipulated in tax treaties is to be applied.

It is also necessary to consider corporate taxes in the third jurisdiction when calculating the strict amount of tax. This is especially true in the case of dividends. However, only a few jurisdictions impose corporate tax on dividend income today[2] (Dittmer 2012). For this reason, corporate taxes in the third jurisdiction are not considered in this study.

**Table 1** 165 jurisdictions subject to this study

| | | | | |
|---|---|---|---|---|
| Afghanistan | Costa Rica (Costa) | Indonesia | Moldova | Senegal |
| Albania | Côte d’Ivoire (Cote) | Iraq | Monaco | Republic of Serbia (Serbia) |
| Algeria | Croatia | Republic of Ireland (Ireland) | Mongolia | Seychelles |
| Angola | Curacao | Isle of Man (Man) | Republic of Montenegro (Montenegro) | Singapore |
| Argentina | Cyprus | Israel | Morocco | Sint Maarten (Sint) |
| Armenia | Czech Republic (Czech) | Italy | Mozambique | Slovak Republic (Slovak) |
| Australia | Denmark | Jamaica | Myanmar | Slovenia |
| Austria | Dominican Republic (DominicanR) | Japan | Namibia | South Africa (South) |
| Azerbaijan | Ecuador | Channel Islands Jersey (Jersey) | Netherlands | South Sudan (Ssudan) |
| Bahamas | Egypt | Jordan | New Zealand (NZ) | Spain |
| Bahrain | El Salvador (El) | Kazakhstan | Nicaragua | Sri Lanka (Sri) |
| Belarus | Equatorial Guinea (Equatorial) | Korea | Nigeria | Suriname |

[2] Many countries have introduced participation exemptions. For example, in Japan, corporate tax is exempted for 95% of dividend income.

| | | | | |
|---|---|---|---|---|
| Belgium | Estonia | Kosovo | Commonwealth of the Northern Mariana Islands (Mariana) | Swaziland |
| Bermuda | Ethiopia | Kuwait | Norway | Sweden |
| Bolivia | Fiji | Kyrgyzstan | Oman | Switzerland |
| Bonaire, Sint Eustatius and Saba (Bonaire) | Finland | Laos | Pakistan | Taiwan |
| Bosnia and Herzegovina (Bosnia) | France | Latvia | Palestinian Authority (Palestinian) | Tanzania |
| Botswana | Gabon | Lebanon | Panama | Thailand |
| Brazil | Georgia | Lesotho | Papua New Guinea (Papua) | Trinidad and Tobago (Trinidad) |
| British Virgin Islands (Virgin) | Germany | Libya | Paraguay | Tunisia |
| Brunei Darussalam (Brunei) | Ghana | Liechtenstein | Peru | Turkey |
| Bulgaria | Gibraltar | Lithuania | Philippines | Uganda |
| Cambodia | Greece | Luxembourg | Poland | Ukraine |
| Cameroon | Guam | Macau Special Administrative Region of China (Macau) | Portugal | the United Arab Emirates (UAE) |
| Canada | Guatemala | Republic of Macedonia (Macedonia) | Puerto Rico (Puerto) | the United Kingdom (UK) |
| Cayman Islands (Cayman) | Channel Islands Guernsey (Guernsey) | Madagascar | Qatar | the United States (US) |
| Cape Verde | Guinea | Malawi | Romania | US Virgin Islands |
| (Cape) | | | | (Uvirgin) |
| Chad | Guyana | Malaysia | Russian Federation (Russian) | Uruguay |
| Chile | Honduras | Maldives | Rwanda | Uzbekistan |
| China | Hong Kong Special Administration Region of China (HK) | Malta | St. Lucia (Lucia) | Venezuela |
| Colombia | Hungary | Mauritania | Saint-Martin (Martin) | Vietnam |
| Democratic Republic of Congo (CongoDR) | Iceland | Mauritius | São Tomé and Príncipe (Sao) | Zambia |
| Republic of Congo (CongoR) | India | Mexico | Saudi Arabia (Saudi) | Zimbabwe |

*Note*: The parentheses represent abbreviations used in the following tables and figures.

## 3.2 Constructing the network

### *3.2.1 Weighted Multi Directed Graph*

As a withholding tax network, we produces a weighted multi directed graph. In the graph, the vertices represent jurisdictions, and all pairs of vertices are connected by arcs because, logically, any jurisdiction can pay dividends, interest, and royalties to any other jurisdictions. Every arc has the withholding tax imposed on dividends, interest, and royalties as a weight. In addition to withholding tax, all arcs also are given a slight weight of $1{\times}10^{-6}$ as a sanction. This is because it costs money, for instance registration fees, to establish a company in a jurisdiction, even if the company is only a paper company.

Therefore, in the case where jurisdiction A imposes a 20% withholding tax on the dividend and jurisdiction B imposes a 30% withholding tax, the dividend's weighted multi directed graph is as shown in Figure 2. In other words, the arc from jurisdiction A to jurisdiction B has $20 + 1{\times}10^{-6}$ as the weight, and the arc from jurisdiction B to jurisdiction A has $30 + 1{\times}10^{-6}$.

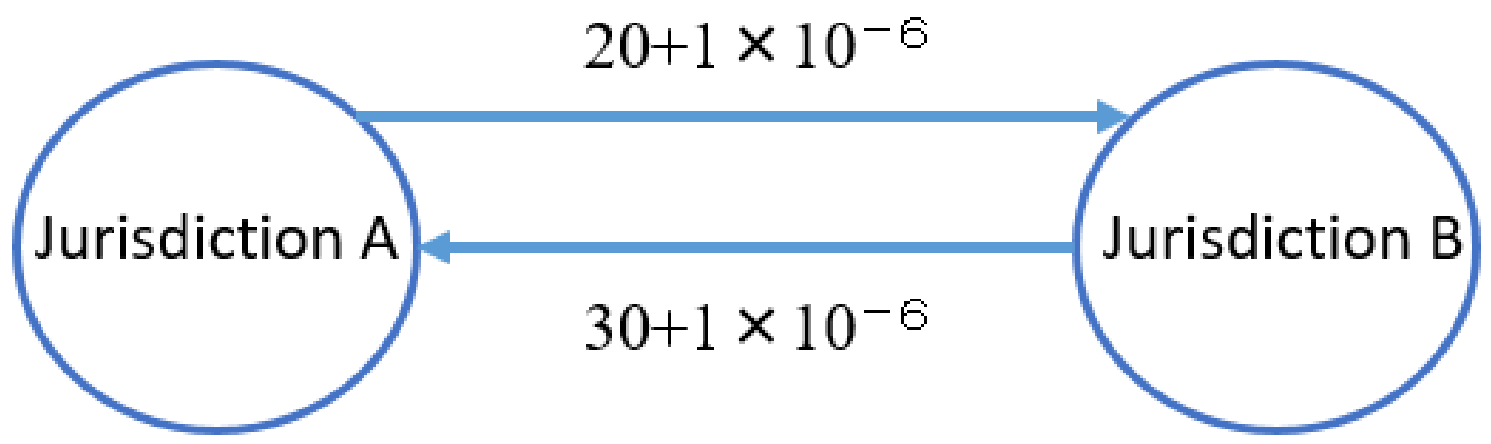

**Fig. 2** An example of jurisdiction A and jurisdiction B

We also created graphs with the arcs removed according to the threshold values. The thresholds provided are seven, 30, 25, 20, 15, 10, 5, and 0, and arcs were removed when weights exceeding the threshold were given. In the case of jurisdiction A and jurisdiction B, the arc from jurisdiction B to jurisdiction A are removed, but in graphs where the threshold value is 25 or less (Figure 3).

**A** **B**

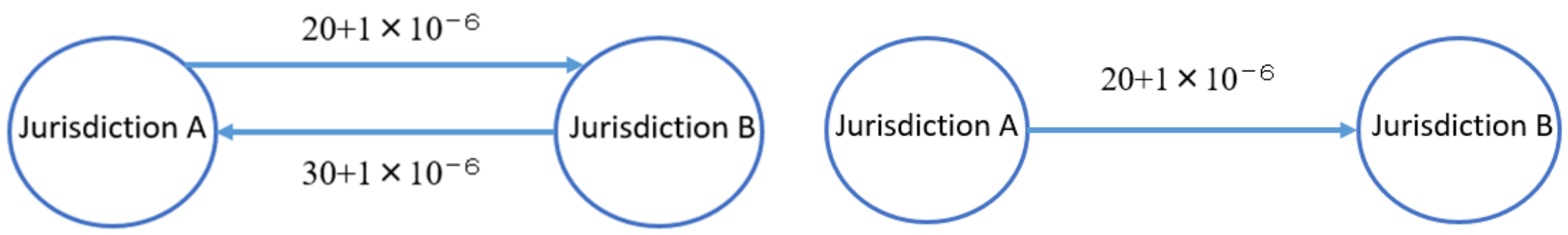

**Fig. 3** An example of jurisdiction A and jurisdiction B. **A**: over threshold of 25, **B**: threshold 25 and 20.

*3.2.2 Weighted Undirected Graph*

We also constructed a network that transformed the weighted multiple directed graphs into weighted undirected graphs. In producing the weighted undirected graphs, the problem we faced is the weight added on each edge, for withholding tax rate differs depending on the jurisdiction paying dividends, interest, and royalties, even if those payments are made between the same two jurisdictions. For example, if a company in the United Kingdom pays dividends to a company in Afghanistan, no withholding tax is imposed on the dividends, but if a company in Afghanistan pays dividends to a company in the United Kingdom, a 20% withholding tax is imposed. The purpose of this paper is to identify which jurisdiction is likely to be used for treaty shopping, so we apply the higher withholding tax rate as a weight. For example, in the case of Afghanistan and the United Kingdom, it looks like Figure 4.

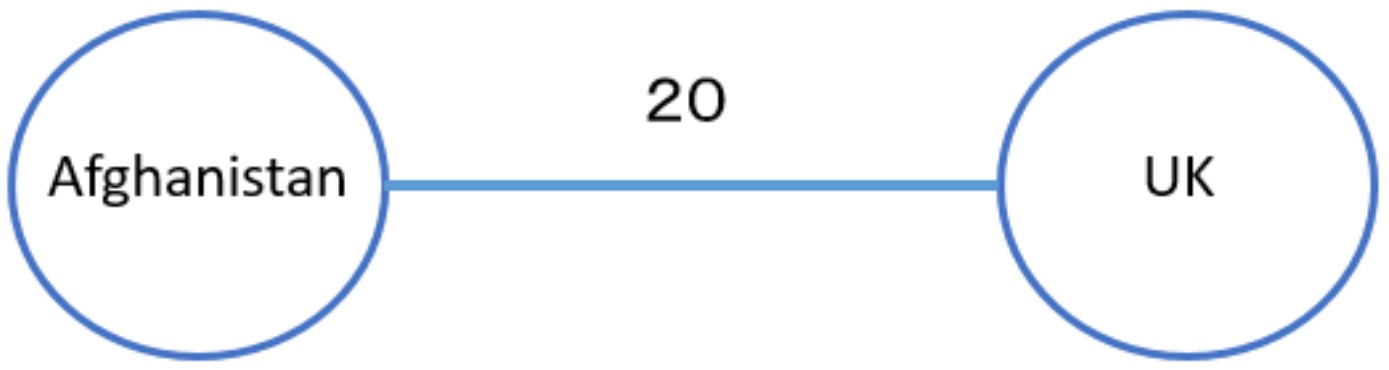

**Fig.4** An example of Afghanistan and the United Kingdom

From the weighted undirected graphs mentioned above, we also created graphs with the edges removed according to the threshold values. This is because we believe it is possible to detect community structures between jurisdictions that uses treaty shopping and the jurisdictions that are used for treaty shopping by removing the edges with high tax rates, that is, the edges that are unlikely to be used for treaty shopping.

The thresholds provided are seven, 30, 25, 20, 15, 10, 5, and 0, and edges were removed when weights exceeding the threshold were given. For example, in the case of Afghanistan and the United Kingdom described above, a weight 20 is given in the weighted undirected graph. Therefore, vertices representing both jurisdictions are connected by edges up to the graph with a threshold value of 20, but in graphs where the threshold value is 15 or less, edges are not connected (Figure 5).

**A** **B**

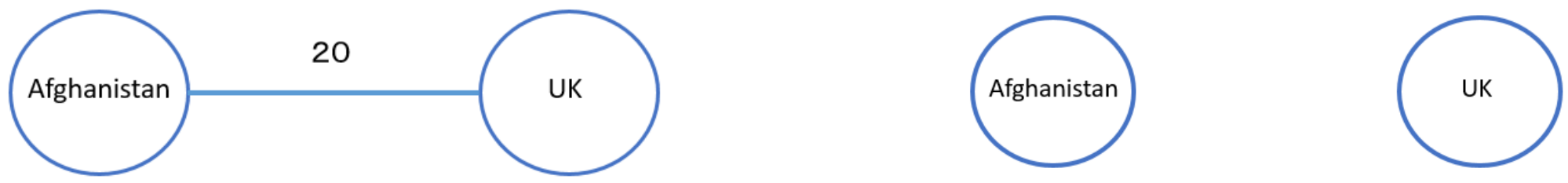

**Fig.5** An example of Afghanistan and the United Kingdom. **A**: over threshold of 20, **B**: under threshold of 15.

## 4 Method of Network Analysis

### 4.1 Centrality calculation

Centrality is an index that shows how important a vertex is for the overall network. The higher the value, the more important the vertex is in the network. However, it is not possible to define unambiguously what are important vertices for the overall network. In this study, centrality is calculated based on the idea that the vertex contributing to the shortest paths in the network is essential to the whole network. This is because the jurisdiction used for treaty shopping is located on the shortest path. By calculating the centrality, we try to estimate which jurisdiction is easier to use for treaty shopping from the viewpoint of withholding tax rates.

We use load centrality as the centrality of the graph (Goh et al. 2001). Load centrality was originally designed to calculate how much data processing capacity each passing point requires for the efficient movement of data packets in a network such as the Internet. The definition is as follows: Suppose one data packet is sent from vertex i to vertex j. The data packet moves along the minimum weight path. When one data packet is sent among all vertices like this, the amount of the data packets which passed through vertex k is represented by $l_k$. The load centrality of vertex k is given by

$$\mathbf{C}_{\boldsymbol{l}}(\mathbf{k}) = \sum_{i \neq j \neq k} \boldsymbol{l}_{i,j}(\boldsymbol{k}) \quad (1)$$

Even though load centrality is sometimes thought to be similar in concept to betweenness centrality (Freeman 1979), the results differ when a graph has multiple minimum weight paths (Brandes 2008). This is because load centrality assumes a data packet is divided equally at the fork of minimum weight paths, while betweenness centrality is divided equally by the number of shortest paths (Figure 6).

In this study, we think of data packets as the amount of dividends, interest, and royalties and withholding tax rates as weight to find the vulnerabilities of the worldwide legal system (1 Introduction).

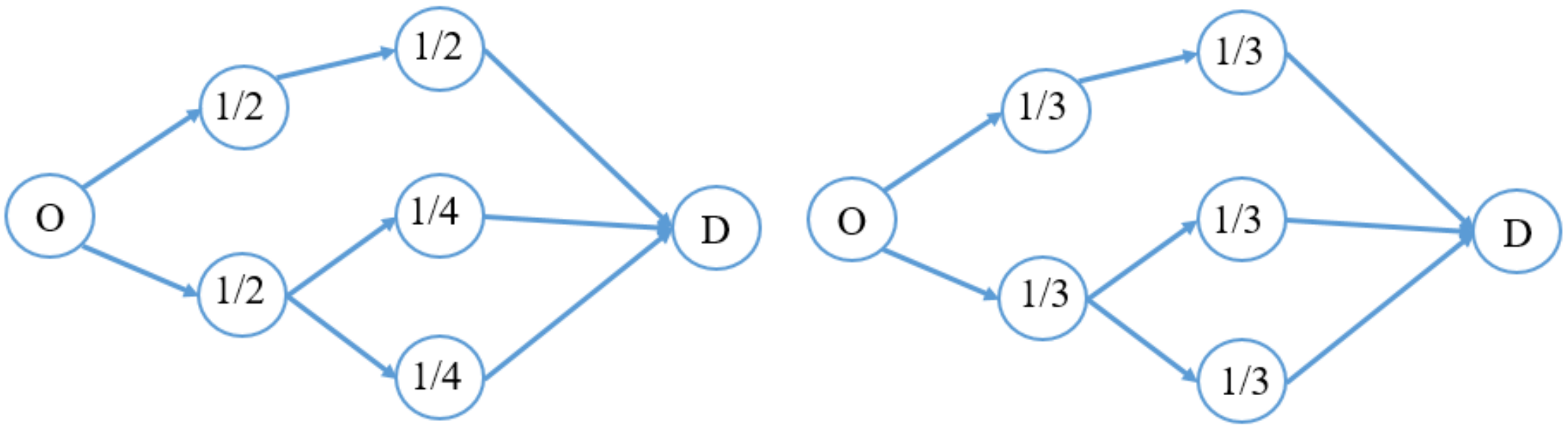

**Fig. 6** Difference in the treatment of multiple shortest paths between load centrality (left) and betweenness centrality (right).

### 4.2 Community detection

A community is a subgraph closely connected in the network. Even though it seems that there is no strict definition of "community" in network science, it is thought that the network has a community structure, if the graph can be divided into subgraphs where the edges connected has density. We detect communities in the withholding tax network because we think that we can find a community structure in the network if there are some relationships between the jurisdictions used for treaty shopping and the jurisdictions having motivations to do treaty shopping.

The problem in detecting communities is whether the communities can be detected appropriately from a network. As we have seen, if the community is regarded as a partial graph closely connected in the network, in evaluating whether there is a community structure in the network, we basically must check whether the network is divided into subgraphs which have many edges in a given subgraph, but few edges between different subgraphs.

We evaluate our results of community detection using modularity, which is often adopted as an index of evaluating community structure. (Newman 2004a; Newman 2004b). To evaluate only significant community structures, Modularity is calculated by subtracting the expected value of the number of edges when considering the community as a random graph $\langle t_{ij} \rangle$ from the number of edges of the divided communities $a_{ij}$:

$$\mathbf{Q} = \frac{1}{2M}\sum_{i,j}^{N}\left(a_{ij} - \frac{k_i k_j}{2M}\right)\delta[C(i), C(j)] \qquad (2)$$

where M is the total number of the edges, $C(i)$ is subset including vertex i, and $\delta[C(i), C(j)]$ is 1 if $C(i)$ and $C(j)$ are the same subset, otherwise it is 0.

In a case of weighted graphs, $a_{ij}$ represents the weight of the edges between vertex i and vertex j rather than the number of edges between them. Similarly, $k_i k_j$ represents the expected weight of the edges assigned randomly to vertex i.

Modularity is normalized so that the maximum value is 1. Therefore, it can be said that there is a strong community structure in the network as it moves closer to 1, but from an empirical point of view, if it is larger than about 0.3, it is often considered that the network has a community structure.

We use the Louvain Method for community detection (Blondel et al. 2008). This is a method which tries to detect communities having high modularity by trying to optimize local and aggregating vertices. Specifically, it is detected as follows: First, the Louvain Method assumes every vertex belongs to one community. Then, calculate how much the value of modularity rises, if each vertex belongs to the same community with an adjacent vertex,

$$\Delta\mathbf{Q} = \left[\frac{\Sigma_{in} + k_{i,in}}{2M} - \left(\frac{\Sigma_{tot} + k_{i,in}}{2M}\right)^2\right] - \left[\frac{\Sigma_{in}}{2M} - \left(\frac{\Sigma_{tot}}{2M}\right)^2 - \left(\frac{k_i}{2M}\right)^2\right] \qquad (3)$$

where, if vertex i belongs to a community, $\boldsymbol{\Sigma_{in}}$ represents the sum of the weights of edges in the community, $\Sigma_{tot}$ represents the sum of the weights of all the edges adjacent to the vertices in the community, $k_{i,in}$ represents the sum of the weights of edges whose vertex i is an end point, $k_i$ represents the sum of the weights of edges in the community whose vertices i are endpoints. The combination of vertices with the greatest modularity rise are specified (the first step). Next, combinations of the specified vertices in the first step are aggregated into one vertex, and the size of the network is reduced (the second step). The first stage and the second stage are recursively repeated, and when the modularity converges, the community is detected. The characteristic of the Louvain Method is not to assume all vertices belong to one community, but to assume each belongs to one community.

## 5 Result

### 5.1 The value of centrality

The top countries differ markedly depending on dividends, interest and royalties and the value of centrality of interest is much lower than that of interest (Table2-4). In the area of international taxation, the Netherlands, Barbados, Cyprus, Estonia, Hungary, Luxembourg, Malta, and Switzerland, and the United Kingdom are cited as examples of jurisdictions commonly known to be likely to be used for treaty shopping (Diamond 2017b). According to the results of this study, we can see that these jurisdictions are the top jurisdictions for dividends.

Regarding the influence of the threshold, although the value of the centrality in dividends is not particularly great (Figure 7), both of the centrality values in interest and royalties are starting to decline from the time when the threshold falls below 10 (Figure 8 and 9). The results may mean that polarization is occurring at withholding tax rates on dividends at above and below 5%, but 10% and 5% withholding tax seems to be the mainstream in many jurisdictions when it comes on interest and royalties.

The results presented in this paper only mean that the tax is less if a company pays its dividends, interest, and royalties through a given jurisdiction, compared to paying them directly to other jurisdictions. It does not mean that such jurisdictions are used for treaty shopping in fact. Some ambivalent tax treaties include provisions for preventing treaty abuse while they offer tax treaty benefits (Okamura 1997). For example, to prevent treaties abuse, jurisdictions such as the United States introduce a limitation on benefits clause, and jurisdictions such as the United Kingdom introduces a principal purpose test. In this study, the presence or absence of these provisions is not considered.

**Table 2** Ranking of the value of load centrality (dividends)

| rank | jurisdiction | centrality | rank | jurisdiction | centrality | rank | jurisdiction | centrality |
|---|---|---|---|---|---|---|---|---|
| 1 | UK | 0.070639 | 30 | Colombia | 0.006188 | 59 | Bahamas | 0.000521 |
| 2 | UAE | 0.058244 | 31 | Zambia | 0.006114 | 60 | Virgin | 0.000521 |
| 3 | Kuwait | 0.044943 | 32 | Tunisia | 0.006075 | 61 | Gibraltar | 0.000521 |
| 4 | Netherlands | 0.027664 | 33 | Germany | 0.005091 | 62 | Guernsey | 0.000521 |
| 5 | Cyprus | 0.024464 | 34 | Liechtenstein | 0.004122 | 63 | Maldives | 0.000521 |
| 6 | HK | 0.0239 | 35 | Latvia | 0.003342 | 64 | Monaco | 0.000521 |
| 7 | Singapore | 0.0228 | 36 | Libya | 0.003185 | 65 | Sao | 0.000521 |
| 8 | Switzerland | 0.022725 | 37 | Bosnia | 0.003125 | 66 | Slovak | 0.000519 |
| 9 | Mauritius | 0.022512 | 38 | Kosovo | 0.002884 | 67 | Macau | 0.000519 |
| 10 | Spain | 0.02209 | 39 | Sweden | 0.002682 | 68 | Cayman | 0.00051 |
| 11 | Luxembourg | 0.020043 | 40 | Myanmar | 0.002109 | 69 | China | 0.000456 |
| 12 | Lucia | 0.018178 | 41 | Brunei | 0.001908 | 70 | Ecuador | 0.000415 |
| 13 | Bahrain | 0.018153 | 42 | Jersey | 0.001433 | 71 | Senegal | 0.000379 |
| 14 | Malaysia | 0.018115 | 43 | Sint | 0.001433 | 72 | Australia | 0.000372 |
| 15 | Qatar | 0.017946 | 44 | Curacao | 0.001422 | 73 | Algeria | 0.000371 |
| 16 | Ireland | 0.017523 | 45 | Bermuda | 0.001305 | 74 | Seychelles | 0.000336 |
| 17 | Estonia | 0.016039 | 46 | Man | 0.001305 | 75 | Jordan | 0.000317 |
| 18 | Malta | 0.015731 | 47 | Norway | 0.001206 | 76 | Czech | 0.000293 |
| 19 | US | 0.01268 | 48 | Vietnam | 0.00114 | 77 | Macedonia | 0.000278 |
| 20 | South | 0.011743 | 49 | Finland | 0.001128 | 78 | Croatia | 0.000256 |
| 21 | Mexico | 0.010967 | 50 | India | 0.001058 | 79 | Romania | 0.000252 |
| 22 | Oman | 0.010475 | 51 | Japan | 0.000949 | 80 | Russian | 0.000244 |
| 23 | Denmark | 0.010308 | 52 | Austria | 0.000829 | 81 | Saudi | 0.000223 |
| 24 | Hungary | 0.010111 | 53 | Iraq | 0.000723 | 82 | Poland | 0.000198 |
| 25 | Belgium | 0.009487 | 54 | Martin | 0.000723 | 83 | Greece | 0.000188 |
| 26 | France | 0.009139 | 55 | Brazil | 0.000706 | 84 | Kyrgyzstan | 0.00017 |
| 27 | Lithuania | 0.007719 | 56 | Cape | 0.000692 | 85 | NZ | 0.00014 |
| 28 | Georgia | 0.007616 | 57 | Madagascar | 0.00068 | 86 | Trinidad | 0.000122 |
| 29 | Bulgaria | 0.006243 | 58 | Palestinian | 0.000657 | | | |

**A**

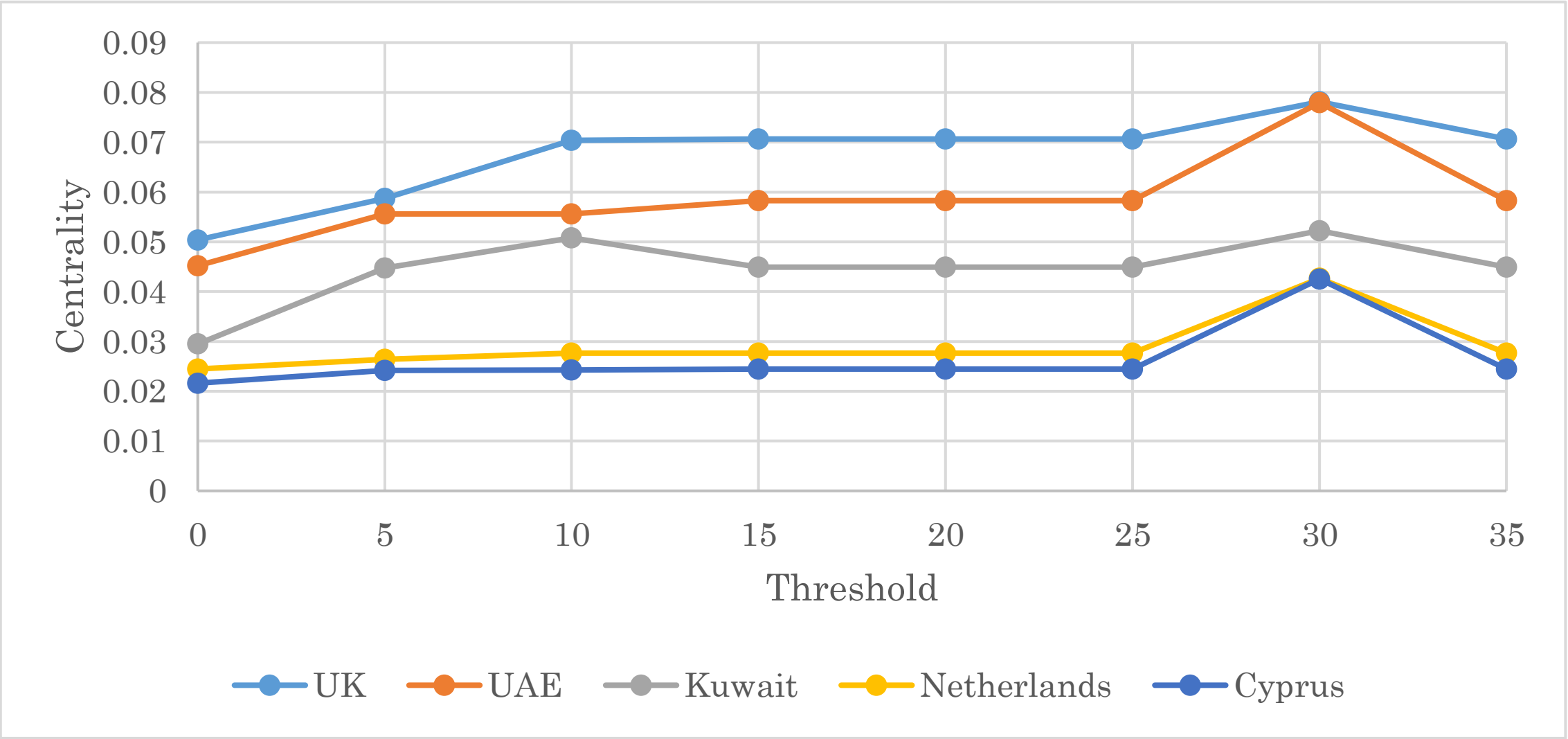

**B**

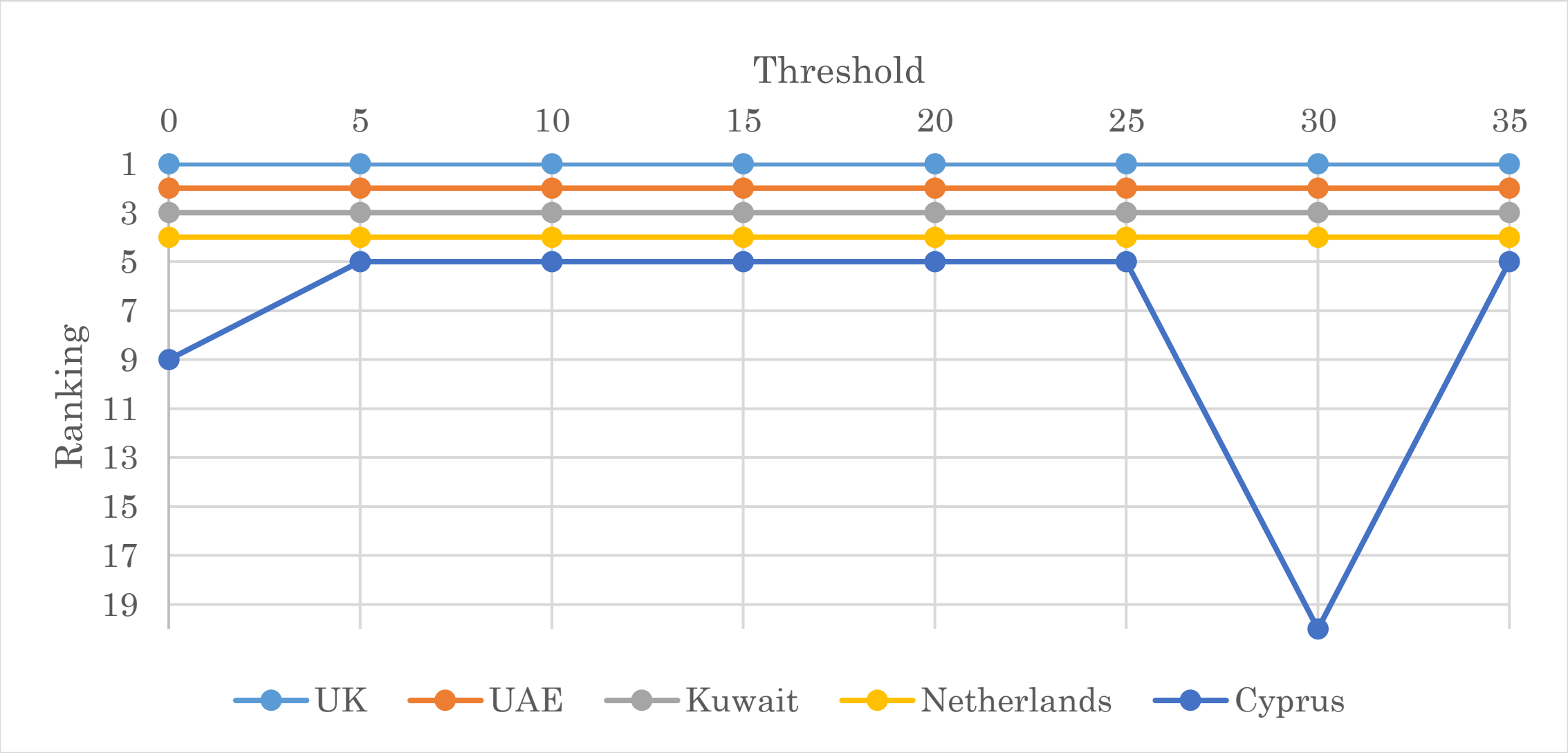

**Fig. 7** **A**: Changes in the top 5 jurisdictions' values of centrality (in the case of dividends). **B**: Change in top the 5 jurisdictions' centrality ranking (in the case of dividends).

**Table 3** Ranking of the value of load centrality (interest)

| rank | jurisdiction | centrality | rank | jurisdiction | centrality | rank | jurisdiction | centrality |
|---|---|---|---|---|---|---|---|---|
| 1 | UAE | 0.041796 | 30 | Israel | 0.006243 | 59 | Gibraltar | 0.000735 |
| 2 | Switzerland | 0.039489 | 31 | Algeria | 0.006228 | 60 | Jersey | 0.000735 |
| 3 | Germany | 0.035639 | 32 | Japan | 0.006183 | 61 | Macau | 0.000735 |
| 4 | France | 0.033565 | 33 | Seychelles | 0.006154 | 62 | Maldives | 0.000735 |
| 5 | UK | 0.032811 | 34 | Portugal | 0.006108 | 63 | Nicaragua | 0.000735 |
| 6 | Hungary | 0.025905 | 35 | Zambia | 0.006101 | 64 | Paraguay | 0.000735 |
| 7 | Canada | 0.024145 | 36 | Cameroon | 0.006067 | 65 | Puerto | 0.000735 |
| 8 | Netherlands | 0.023775 | 37 | Denmark | 0.005954 | 66 | Martin | 0.000735 |
| 9 | Sweden | 0.023619 | 38 | HK | 0.003959 | 67 | Sao | 0.000735 |
| 10 | Luxembourg | 0.022488 | 39 | Finland | 0.003613 | 68 | Sint | 0.000735 |
| 11 | Kuwait | 0.019532 | 40 | Bermuda | 0.003085 | 69 | Belgium | 0.000732 |
| 12 | Norway | 0.016801 | 41 | Man | 0.003085 | 70 | Cayman | 0.000724 |
| 13 | Ireland | 0.016736 | 42 | Poland | 0.003044 | 71 | Curacao | 0.000724 |
| 14 | Czech | 0.016253 | 43 | Slovak | 0.001952 | 72 | Singapore | 0.000533 |
| 15 | US | 0.014057 | 44 | Guernsey | 0.001901 | 73 | Macedonia | 0.000456 |
| 16 | Estonia | 0.013082 | 45 | Monaco | 0.001901 | 74 | Australia | 0.000344 |
| 17 | Malta | 0.011142 | 46 | Spain | 0.001657 | 75 | Croatia | 0.000321 |
| 18 | Austria | 0.01061 | 47 | Indonesia | 0.001499 | 76 | Romania | 0.000262 |
| 19 | South | 0.010495 | 48 | Colombia | 0.001325 | 77 | Korea | 0.000247 |
| 20 | Cyprus | 0.009644 | 49 | Libya | 0.001266 | 78 | Saudi | 0.000242 |
| 21 | Russian | 0.009492 | 50 | Liechtenstein | 0.00126 | 79 | Greece | 0.000236 |
| 22 | Bahrain | 0.009354 | 51 | Malawi | 0.00119 | 80 | Panama | 0.000211 |
| 23 | Latvia | 0.009251 | 52 | Zimbabwe | 0.001053 | 81 | Iceland | 0.000121 |
| 24 | Italy | 0.008287 | 53 | Qatar | 0.00104 | | | |
| 25 | Belarus | 0.006977 | 54 | Suriname | 0.000991 | | | |
| 26 | Mauritius | 0.006841 | 55 | Slovenia | 0.0009 | | | |
| 27 | Oman | 0.006553 | 56 | Georgia | 0.000888 | | | |
| 28 | Ukraine | 0.006507 | 57 | Bahamas | 0.000735 | | | |
| 29 | Bulgaria | 0.006357 | 58 | Virgin | 0.000735 | | | |

**A**

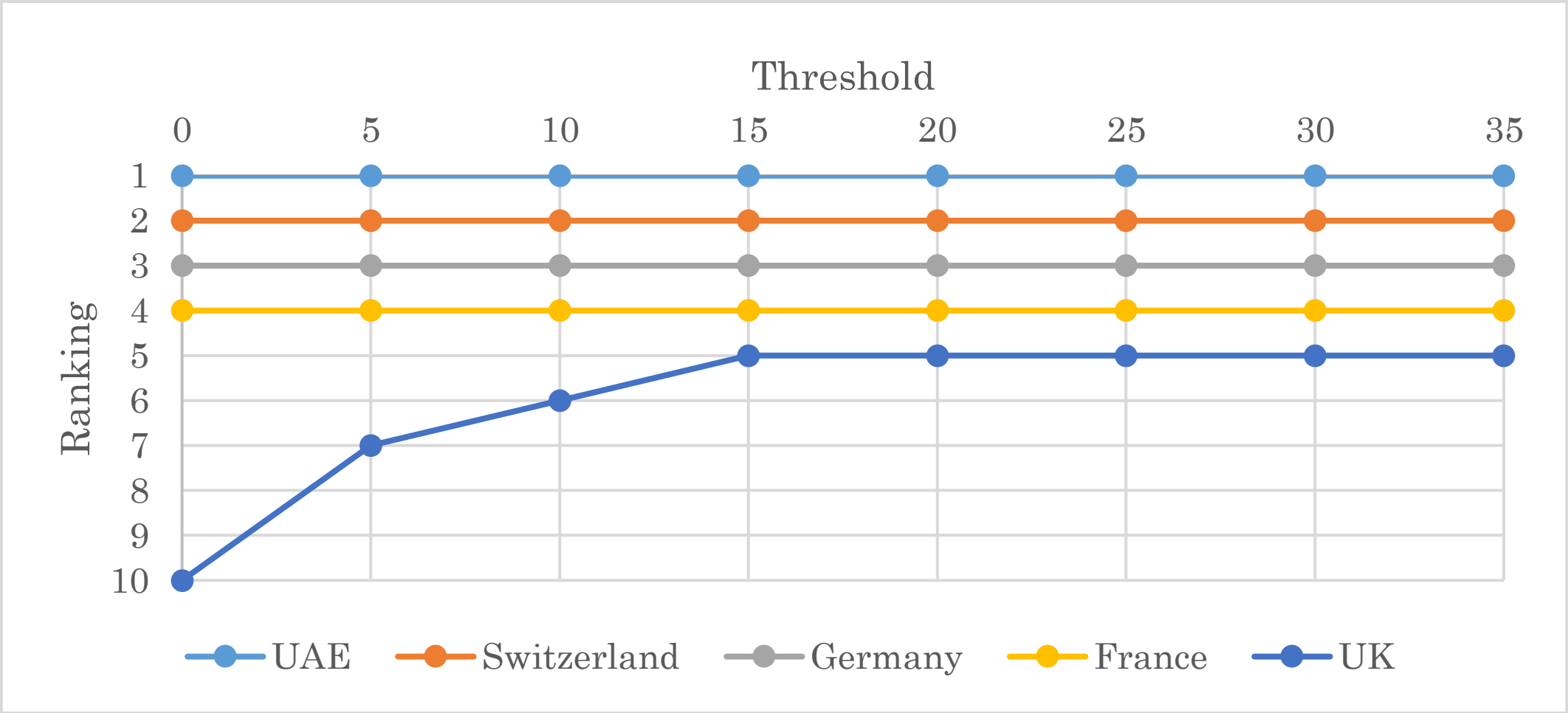

**B**

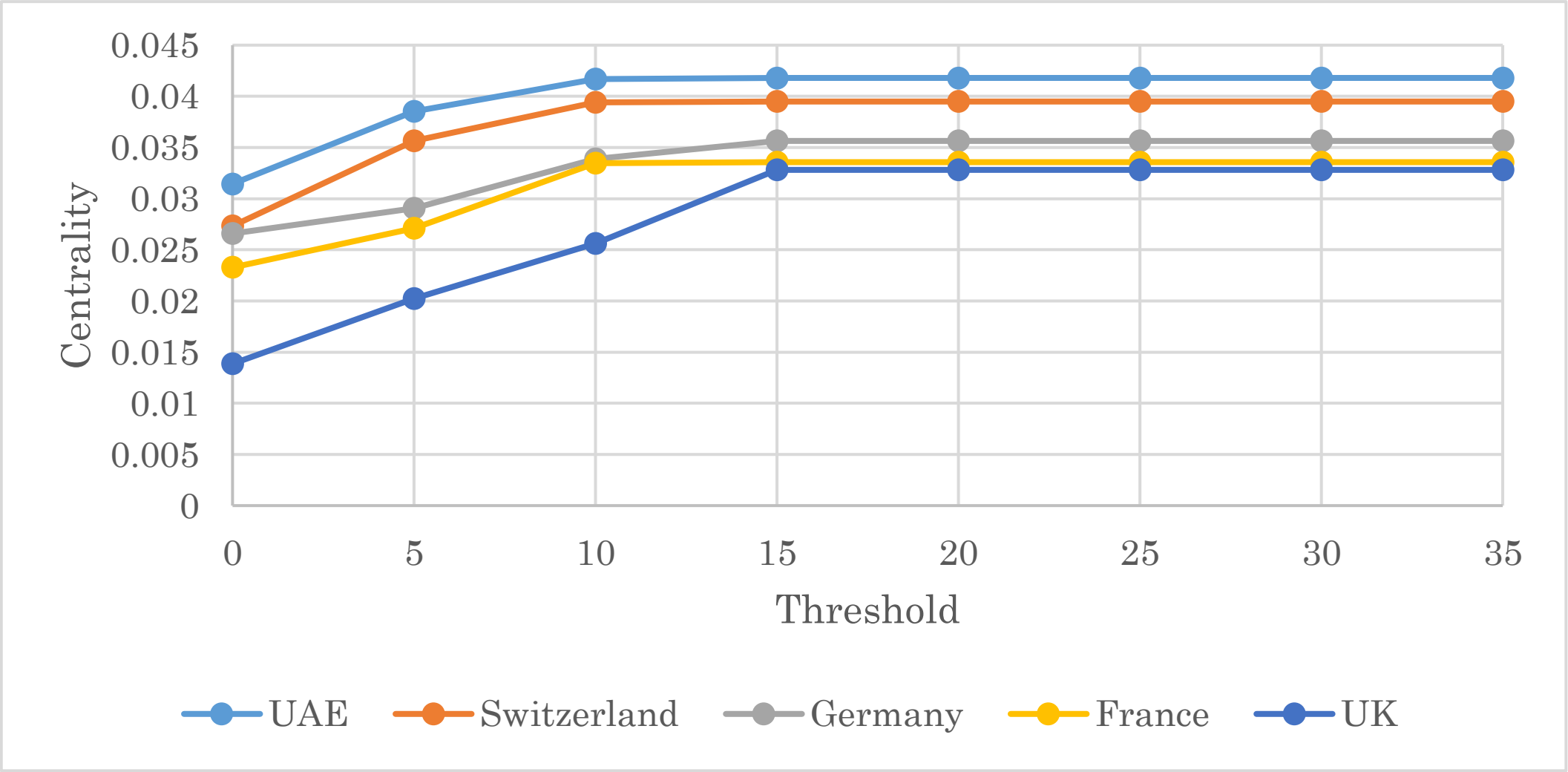

**Fig. 8 A**: Changes in the top 5 jurisdictions' values of centrality (in the case of interest). **B**: Changes in the top 5 jurisdictions' centrality (in the case of interest).

**Table 4** Ranking of the value of load centrality (royalties)

| rank | jurisdiction | centrality | rank | jurisdiction | centrality | rank | jurisdiction | centrality |
|---|---|---|---|---|---|---|---|---|
| 1 | UAE | 0.079455 | 30 | Estonia | 0.007717 | 59 | Bulgaria | 0.000396 |
| 2 | Switzerland | 0.079043 | 31 | Slovak | 0.006905 | 60 | HK | 0.000318 |
| 3 | France | 0.05139 | 32 | Austria | 0.006784 | 61 | Bahamas | 0.000312 |
| 4 | Hungary | 0.048268 | 33 | Zambia | 0.006121 | 62 | Virgin | 0.000312 |
| 5 | Mauritius | 0.047201 | 34 | Libya | 0.005969 | 63 | Gibraltar | 0.000312 |
| 6 | Sweden | 0.045731 | 35 | Liechtenstein | 0.005768 | 64 | Sao | 0.000312 |
| 7 | Netherlands | 0.042169 | 36 | Belgium | 0.005208 | 65 | Sint | 0.000312 |
| 8 | Norway | 0.037786 | 37 | Russian | 0.005146 | 66 | Cayman | 0.000303 |
| 9 | Cyprus | 0.034533 | 38 | Bermuda | 0.003427 | 67 | Curacao | 0.000303 |
| 10 | Ireland | 0.029864 | 39 | Kuwait | 0.003319 | 68 | Iceland | 0.000196 |
| 11 | Luxembourg | 0.029044 | 40 | Finland | 0.002753 | 69 | Cameroon | 0.000164 |
| 12 | Senegal | 0.029034 | 41 | Georgia | 0.001943 | 70 | Belarus | 0.00014 |
| 13 | UK | 0.024726 | 42 | Nicaragua | 0.001733 | 71 | Malaysia | 0.000116 |
| 14 | Malta | 0.02437 | 43 | Paraguay | 0.001733 | | | |
| 15 | Gabon | 0.024012 | 44 | Puerto | 0.001733 | | | |
| 16 | US | 0.019653 | 45 | Martin | 0.001733 | | | |
| 17 | Spain | 0.018766 | 46 | Suriname | 0.001733 | | | |
| 18 | Bahrain | 0.018091 | 47 | Tunisia | 0.001673 | | | |
| 19 | Latvia | 0.01757 | 48 | Man | 0.001374 | | | |
| 20 | Germany | 0.016731 | 49 | Jersey | 0.001374 | | | |
| 21 | Macau | 0.01241 | 50 | Ukraine | 0.001295 | | | |
| 22 | Canada | 0.012097 | 51 | China | 0.001124 | | | |
| 23 | Monaco | 0.01029 | 52 | Czech | 0.001086 | | | |
| 24 | Korea | 0.009735 | 53 | Greece | 0.001037 | | | |
| 25 | South | 0.009731 | 54 | Singapore | 0.000906 | | | |
| 26 | Japan | 0.008836 | 55 | Croatia | 0.000821 | | | |
| 27 | Denmark | 0.007992 | 56 | Cote | 0.000541 | | | |
| 28 | Italy | 0.007951 | 57 | Seychelles | 0.00047 | | | |
| 29 | Guernsey | 0.007783 | 58 | Israel | 0.000448 | | | |

**A**

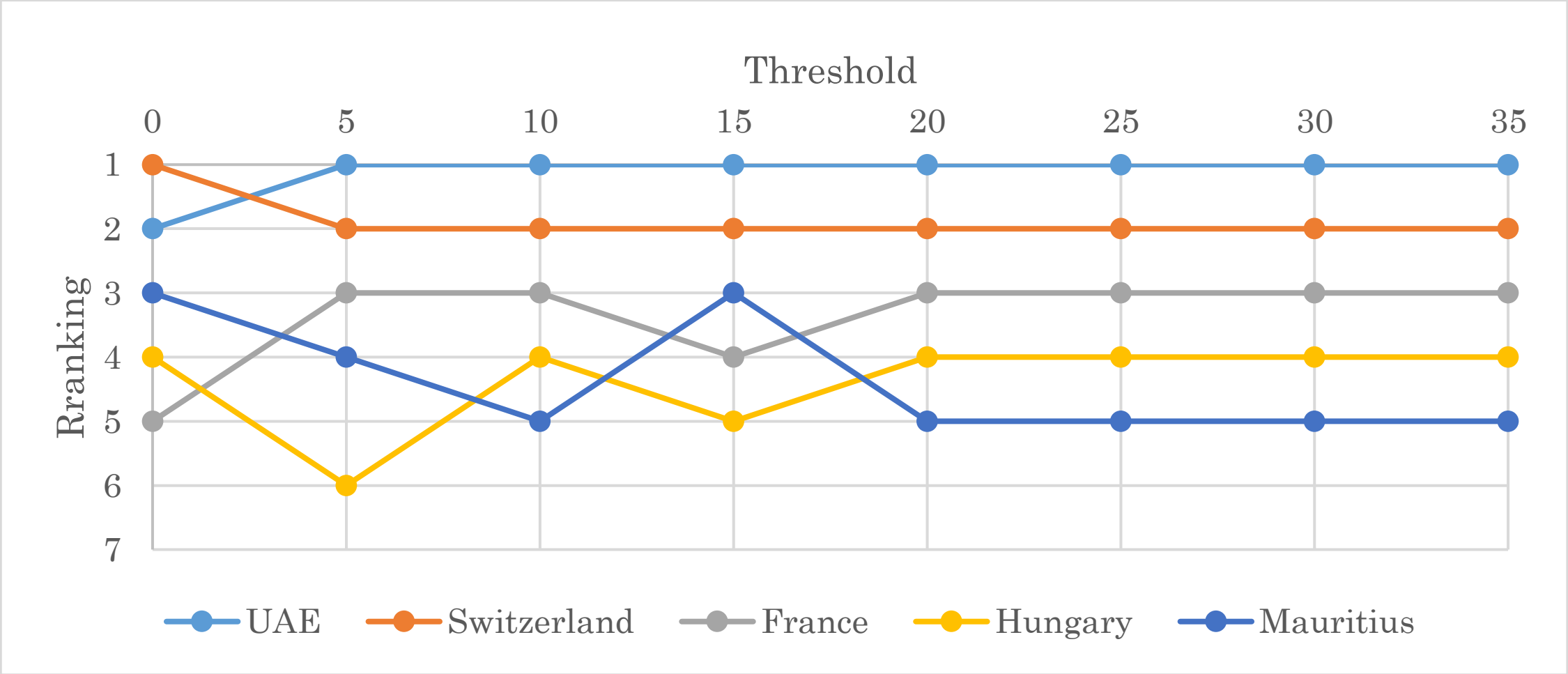

**B**

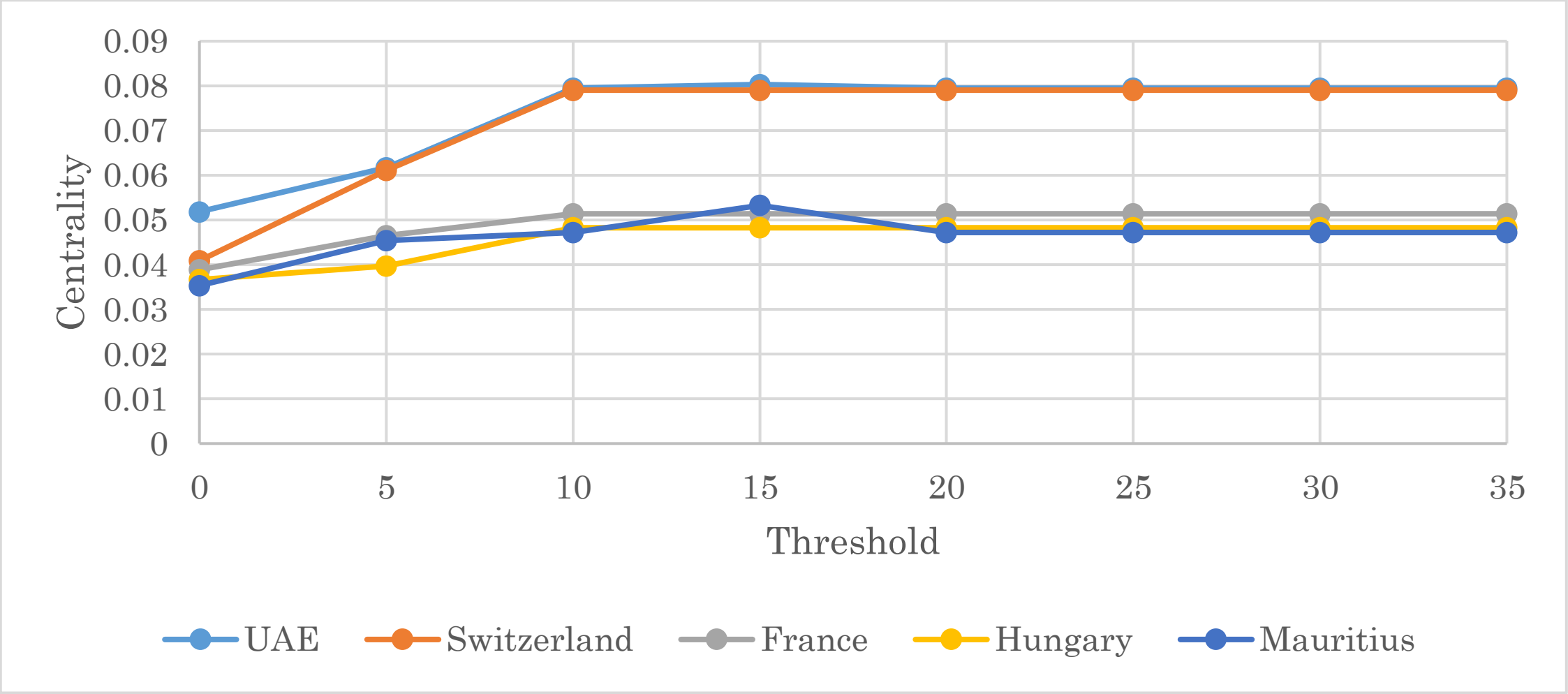

**Fig. 9** **A**: Changes in the top 5 jurisdictions' values of centrality (in the case of royalties). **B**: Changes in the top 5 jurisdictions' centrality ranking (in the case of royalties).

## 5.2 Communities Founded

### *5.2.1 Dividends*

The weighted undirected graph with a threshold of 5% recorded the highest modularity (Figure 10). Its value is 0.2853738, or close to 0.3, so it can be said that the withholding tax network for dividends has a community structure. The results are as shown in Table 5 and jurisdictions in each community are arranged in descending order of centrality, which is in the weighted multi directed graph with a threshold of 5%. We identified four communities. All except community 4 has jurisdictions whose centralities are high. Community 1 includes the United Kingdom, Kuwait, the Netherlands, Switzerland, Spain, Luxembourg,

Ireland, Estonia, Malta, the United States, and other jurisdictions, community 2 includes Cyprus, Hong Kong, Lucia, Bahrain, Mauritius, and others, and community 3 includes the United Arab Emirates, Singapore, Malaysia, Qatar, and other. We cannot find a relationship between the detected communities and each jurisdiction's location. The interesting thing is that mainland China and Hong Kong as well as India and Mauritius, which are known for their compatibility, are not in the same communities.

In the weighted undirected graph with a threshold of 5%, there were 34 vertices with no edges with other vertices. This means that dividends are subject to a withholding tax rate of more than 5% when paid from these the jurisdictions or for the jurisdictions. Therefore, it can be said that the possibility that these jurisdictions will be used for treaty shopping involved dividends is extremely low. However, the possibility that dividends will be paid for these jurisdictions remains, given that of the 165 jurisdictions subject to this analysis, some jurisdictions do not impose withholding tax initially. Therefore, if dividends are paid for 34 jurisdictions without links in a 5% threshold weighted undirected graph, at least these jurisdictions not imposing withholding tax. Among the 34 jurisdictions, it is possible that the U.S. Virgin Islands will be used as a cash box, where corporate profits are accumulated, given their lack of corporate taxation.

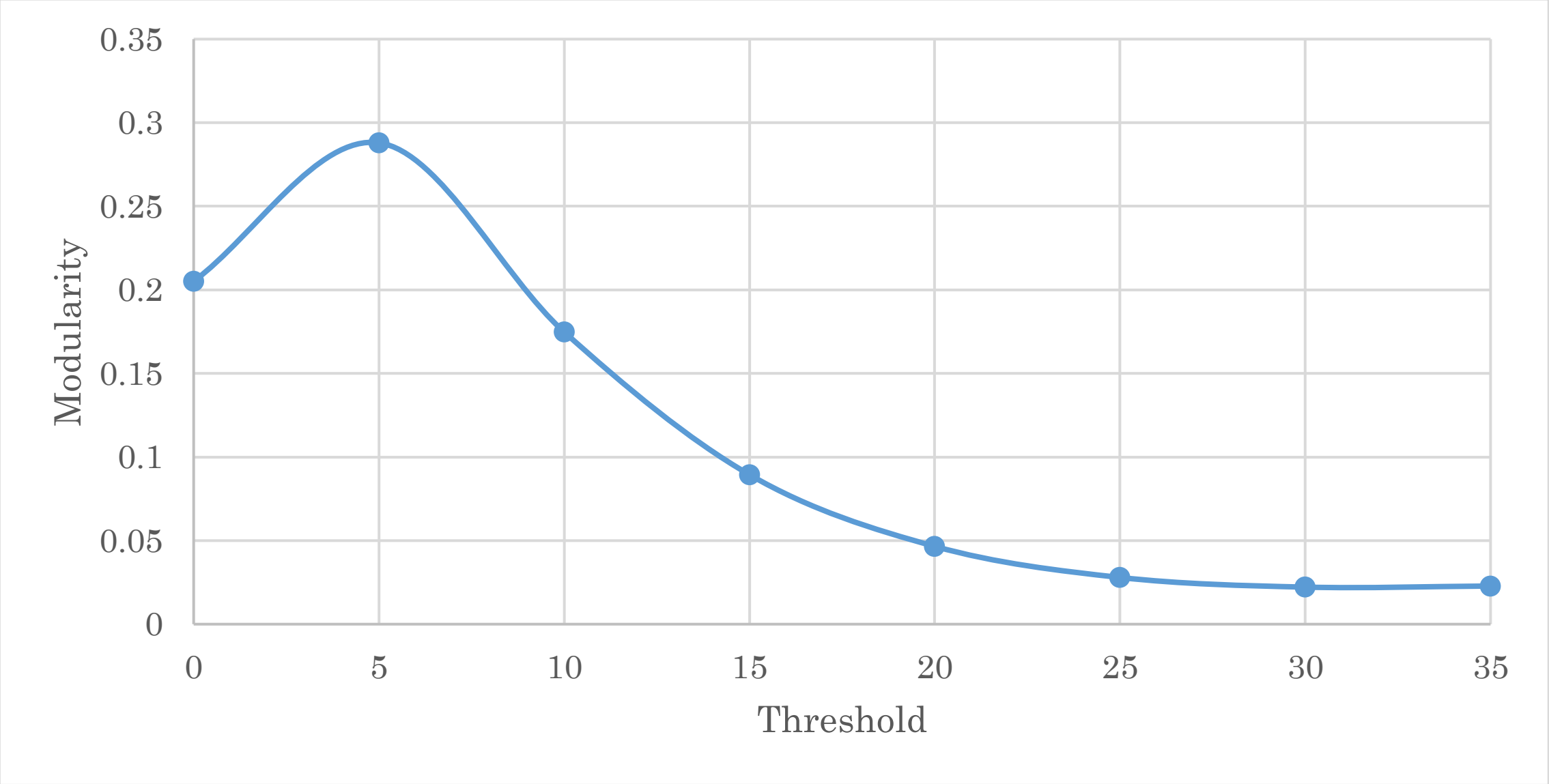

**Fig. 10** Change in modularity, threshold 35 – threshold 0 (in the case of dividends)

**Table 5** Communities (dividends)

| **Community 1** | | **Community 2** | | **Community 3** | | **no kink** | |
|---|---|---|---|---|---|---|---|
| jurisdiction | centrality | jurisdiction | centrality | jurisdiction | centrality | jurisdiction | centrality |
| UK | 0.05873 | Cyprus | 0.02422 | UAE | 0.05556 | Mariana | 5.39E-06 |
| Kuwait | 0.04473 | HK | 0.02361 | Singapore | 0.02252 | Tanzania | 8.31E-07 |
| Netherlands | 0.02641 | Lucia | 0.01814 | Malaysia | 0.01811 | Afghanistan | 0 |
| Switzerland | 0.02246 | Bahrain | 0.01786 | Qatar | 0.01596 | Angola | 0 |
| Spain | 0.02187 | Mauritius | 0.01203 | Norway | 0.0012 | Argentina | 0 |
| Luxembourg | 0.01975 | Oman | 0.01018 | Vietnam | 0.00113 | Bolivia | 0 |
| Ireland | 0.01744 | Georgia | 0.0076 | India | 0.00104 | Cambodia | 0 |
| Estonia | 0.01575 | Bulgaria | 0.00624 | Japan | 0.00094 | Cameroon | 0 |
| Malta | 0.0155 | Colombia | 0.00619 | NZ | 0.00014 | Chad | 0 |
| US | 0.01268 | Liechtenstein | 0.00409 | Venezuela | 1.95E-05 | CongoDR | 0 |
| South | 0.01173 | Libya | 0.00315 | Fiji | 5.67E-06 | CongoR | 0 |
| Mexico | 0.01096 | Kosovo | 0.00285 | Mozambique | 5.67E-06 | DominicanR | 0 |
| Denmark | 0.01016 | Myanmar | 0.00207 | Uruguay | 4.96E-06 | Equatorial | 0 |
| Hungary | 0.00983 | Brunei | 0.00187 | Portugal | 1.61E-06 | Ethiopia | 0 |
| Lithuania | 0.00749 | Jersey | 0.0014 | Azerbaijan | 0 | Gabon | 0 |
| France | 0.00697 | Sint | 0.0014 | Laos | 0 | Guam | 0 |
| Zambia | 0.00611 | Curacao | 0.00139 | Morocco | 0 | Guinea | 0 |
| Belgium | 0.00545 | Bermuda | 0.00127 | Namibia | 0 | Honduras | 0 |
| Germany | 0.00508 | Man | 0.00127 | Pakistan | 0 | Lebanon | 0 |
| Latvia | 0.00318 | Iraq | 0.00069 | | | Lesotho | 0 |
| Bosnia | 0.0031 | Martin | 0.00069 | **Community 4** | | Malawi | 0 |
| Sweden | 0.00253 | Brazil | 0.00068 | jurisdiction | centrality | Mauritania | 0 |
| Finland | 0.00105 | Madagascar | 0.00067 | Seychelles | 0.00011 | Nicaragua | 0 |
| Austria | 0.00083 | Cape | 0.00066 | Zimbabwe | 4.28E-05 | Nigeria | 0 |
| China | 0.00045 | Palestinian | 0.00062 | Botswana | 0 | Papua | 0 |
| Slovak | 0.00039 | Bahamas | 0.00049 | | | Paraguay | 0 |
| Algeria | 0.00037 | Virgin | 0.00049 | | | Philippines | 0 |
| Australia | 0.0003 | Gibraltar | 0.00049 | | | Puerto | 0 |
| Macedonia | 0.00027 | Guernsey | 0.00049 | | | Rwanda | 0 |
| Croatia | 0.00025 | Maldives | 0.00049 | | | Ssudan | 0 |
| Poland | 0.00019 | Monaco | 0.00049 | | | Suriname | 0 |
| Czech | 0.00019 | Sao | 0.00049 | | | Swaziland | 0 |
| Russian | 0.00019 | Macau | 0.00048 | | | Thailand | 0 |
| Greece | 0.00018 | Cayman | 0.00048 | | | Uvirgin | 0 |
| Kyrgyzstan | 0.00017 | Ecuador | 0.00039 | | | | |
| Belarus | 7.86E-05 | Jordan | 0.0003 | | | | |
| Uzbekistan | 7.60E-05 | Saudi | 0.00022 | | | | |
| Albania | 7.42E-05 | Romania | 0.00018 | | | | |
| Armenia | 5.75E-05 | Trinidad | 0.00012 | | | | |
| Slovenia | 4.65E-05 | Tunisia | 1.53E-05 | | | | |
| Egypt | 3.12E-05 | El | 7.70E-06 | | | | |
| Panama | 3.01E-05 | Guyana | 6.46E-06 | | | | |
| Ukraine | 2.19E-05 | Bonaire | 0 | | | | |
| Mongolia | 1.67E-05 | Cote | 0 | | | | |
| Korea | 1.57E-05 | Guatemala | 0 | | | | |
| Iceland | 1.29E-05 | Indonesia | 0 | | | | |
| Moldova | 1.29E-05 | Peru | 0 | | | | |
| Israel | 1.14E-05 | Sri | 0 | | | | |
| Chile | 9.80E-06 | | | | | | |
| Canada | 8.15E-06 | | | | | | |
| Uganda | 6.15E-06 | | | | | | |
| Senegal | 4.48E-06 | | | | | | |
| Jamaica | 2.72E-06 | | | | | | |
| Italy | 2.20E-06 | | | | | | |
| Costa | 0 | | | | | | |
| Ghana | 0 | | | | | | |
| Kazakhstan | 0 | | | | | | |
| Montenegro | 0 | | | | | | |
| Serbia | 0 | | | | | | |
| Taiwan | 0 | | | | | | |
| Turkey | 0 | | | | | | |

*5.2.2 Interest*

The weighted undirected network with a threshold of 5% recorded the highest modularity (Figure 11). Its value is 0.283715, or close to 0.3, so it can be said that the withholding tax network for interest has a community structure. The results are as shown in Table 6 and jurisdictions in each community are arranged in descending order of centrality, which is in the weighted multi directed graph with a threshold of 5%. We identified four communities. All except community 4 have jurisdictions whose centralities are somewhat high. Community 1 includes Switzerland, Germany, France, Kuwait, the United Kingdom, Ireland, and other jurisdictions, community 2 includes Hungary, the Netherlands, Sweden, and others, and community 3 includes the United Arab Emirates and other. Although the United Arab Emirates has a high centrality value (Table 3), the size of the community to which it belongs is small (Table 6). Therefore, it may not be suitable for treaty shopping. Many European countries belong to Community 1 or 2. This may be because the European Union has issued an "Interest Directive" (2003/49/EC), which exempts corporations from paying withholding tax on interest paid within the EU member jurisdictions, to help create EU single market. There were 47 vertices not having edges with other vertices. Regarding these vertices, the same thing can be said as for the dividends (5.2.1 Dividends).

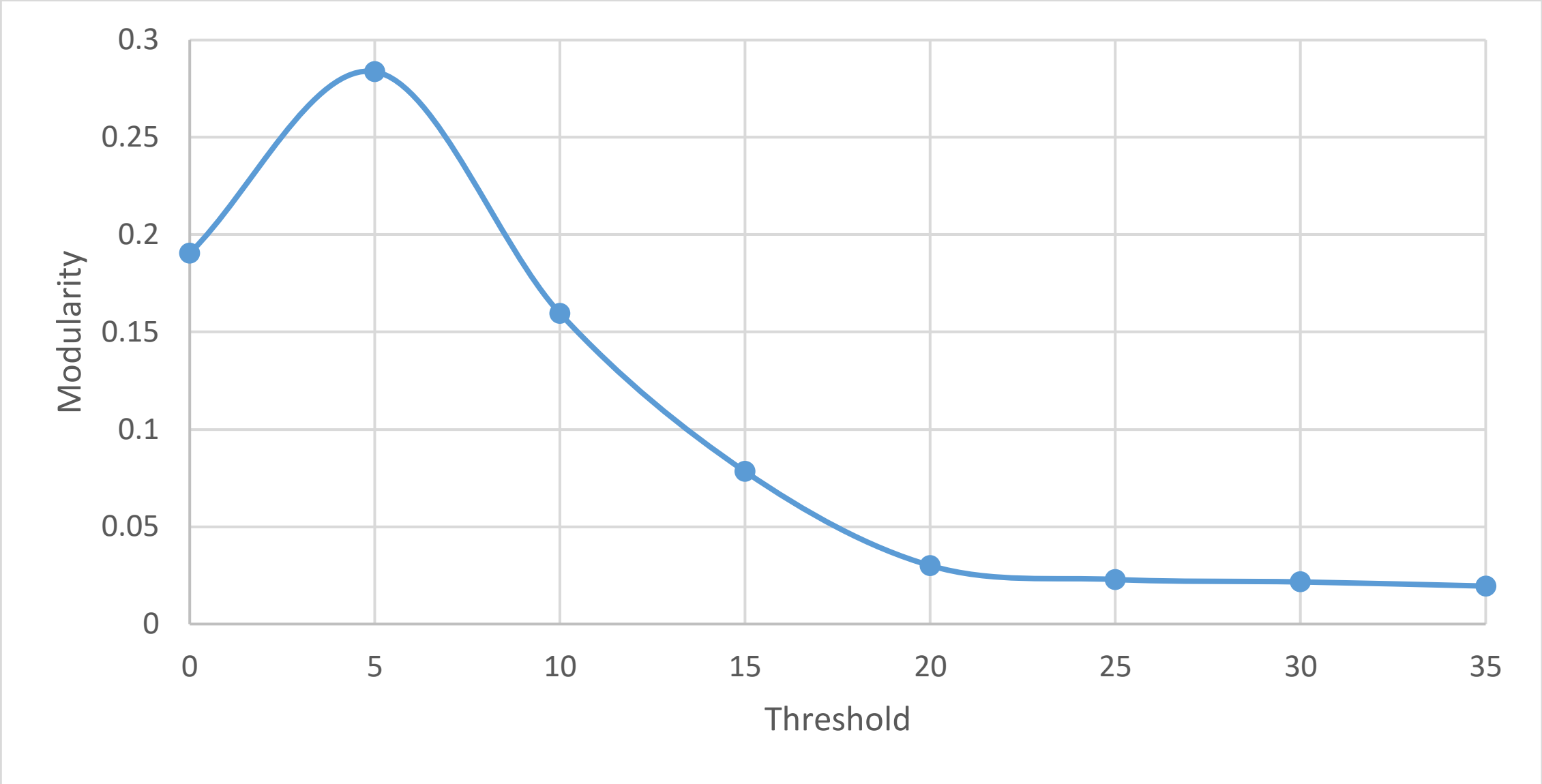

**Fig. 11** Change in modularity, threshold 35 – threshold 0 (in the case of interest)

**Table 6** Communities (interest)

| **Community 1** | | **Community 2** | | **Community 3** | | **no link** | |
|---|---|---|---|---|---|---|---|
| jurisdiction | centrality | jurisdiction | centrality | jurisdiction | centrality | jurisdiction | centrality |
| Switzerland | 0.035669 | Hungary | 0.022507 | UAE | 0.038503 | Afghanistan | 0 |
| Germany | 0.029059 | Netherlands | 0.018449 | Bahrain | 0.008041 | Angola | 0 |
| France | 0.027107 | Sweden | 0.018065 | Mauritius | 0.005814 | Bolivia | 0 |
| Kuwait | 0.022897 | Canada | 0.015561 | Oman | 0.00555 | Botswana | 0 |
| UK | 0.020226 | Luxembourg | 0.013706 | Belgium | 0.000186 | Brazil | 0 |
| Ireland | 0.016536 | Norway | 0.013524 | Seychelles | 4.37E-05 | Brunei | 0 |
| US | 0.013929 | Portugal | 0.006106 | Lithuania | 4.25E-05 | Cambodia | 0 |
| Czech | 0.013042 | Cameroon | 0.006054 | Swaziland | 1.04E-05 | Cape | 1.45E-05 |
| Russian | 0.010392 | Estonia | 0.005487 | Mozambique | 4.16E-06 | Chad | 0 |
| Austria | 0.009228 | HK | 0.003732 | Senegal | 2.84E-06 | CongoDR | 0 |
| Latvia | 0.008363 | Bermuda | 0.001896 | Fiji | 2.67E-06 | Cote | 0 |
| Malta | 0.008265 | Man | 0.001896 | Malaysia | 0 | DominicanR | 0 |
| Cyprus | 0.008104 | Guernsey | 0.001793 | Tunisia | 0 | El | 0 |
| Italy | 0.007799 | Monaco | 0.001793 | | | Equatorial | 0 |
| Zambia | 0.006101 | Slovak | 0.001587 | **Community 4** | | Gabon | 0 |
| China | 0.006098 | Colombia | 0.001112 | jurisdiction | centrality | Ghana | 0 |
| Ukraine | 0.005536 | Libya | 0.00104 | Algeria | 0.006217 | Guam | 0 |
| Denmark | 0.003865 | Liechtenstein | 0.001033 | Egypt | 0 | Guatemala | 0 |
| Finland | 0.003376 | Suriname | 0.000973 | | | Guinea | 0 |
| Poland | 0.002567 | Malawi | 0.000967 | | | Guyana | 0 |
| South | 0.002509 | Zimbabwe | 0.000827 | | | Honduras | 0 |
| Spain | 0.001382 | Bahamas | 0.000717 | | | India | 0 |
| Qatar | 0.000989 | Virgin | 0.000717 | | | Iraq | 0 |
| Indonesia | 0.000893 | Gibraltar | 0.000717 | | | Jamaica | 0 |
| Belarus | 0.000813 | Jersey | 0.000717 | | | Kazakhstan | 0 |
| Slovenia | 0.000588 | Macau | 0.000717 | | | Laos | 0 |
| Singapore | 0.000468 | Maldives | 0.000717 | | | Lebanon | 0 |
| Macedonia | 0.000358 | Nicaragua | 0.000717 | | | Lesotho | 0 |
| Bulgaria | 0.000216 | Paraguay | 0.000717 | | | Madagascar | 0 |
| Korea | 0.000214 | Puerto | 0.000717 | | | Mauritania | 0 |
| Croatia | 0.00021 | Martin | 0.000717 | | | Morocco | 0 |
| Greece | 0.0002 | Sao | 0.000717 | | | Myanmar | 0 |
| Romania | 0.000134 | Sint | 0.000717 | | | Nigeria | 0 |
| Panama | 0.000134 | Cayman | 0.000705 | | | Mariana | 2.82E-06 |
| Israel | 0.000127 | Curacao | 0.000705 | | | Pakistan | 0 |
| Iceland | 0.0001 | Georgia | 0.000637 | | | Palestinian | 0 |
| Mongolia | 7.51E-05 | Saudi | 0.000226 | | | Papua | 0 |
| Armenia | 6.89E-05 | Australia | 0.000188 | | | Philippines | 0 |
| Japan | 6.57E-05 | Argentina | 5.25E-05 | | | Rwanda | 0 |
| Albania | 4.40E-06 | Ecuador | 2.91E-05 | | | Lucia | 0 |
| Kyrgyzstan | 3.96E-05 | NZ | 1.44E-05 | | | Ssudan | 0 |
| Venezuela | 3.03E-05 | Jordan | 1.30E-05 | | | Taiwan | 0 |
| Bosnia | 2.88E-05 | Bonaire | 0 | | | Tanzania | 4.79E-06 |
| Uruguay | 2.18E-05 | Chile | 0 | | | Thailand | 0 |
| Serbia | 1.66E-05 | Ethiopia | 0 | | | Trinidad | 0 |
| Montenegro | 1.58E-05 | Peru | 0 | | | Uganda | 0 |
| Moldova | 1.44E-05 | Sri | 0 | | | Uvirgin | 0 |
| Kosovo | 1.17E-05 | Vietnam | 0 | | | | |
| Uzbekistan | 1.04E-05 | | | | | | |
| CongoR | 3.00E-06 | | | | | | |

| | | | | |
|---|---|---|---|---|
| Namibia | 2.59E-06 | | | |
| Azerbaijan | 0 | | | |
| Costa | 0 | | | |
| Mexico | 0 | | | |
| Turkey | 0 | | | |

*5.2.3 Royalties*

The weighted undirected graph with a threshold of 5% recorded the highest modularity (Figure 12). Its value is 0.325377, which is over 0.3. Thus, the withholding tax network on royalties has a community structure. The results are as shown in Table 9 and jurisdictions in each community are arranged in descending order of centrality, which is in the weighted multi directed graph with a threshold of 5%. We identified five communities. All except community 5 have jurisdictions whose centralities are high. Community 1 includes Switzerland, Hungary, the Netherlands, Ireland, Malta, and so on, community 2 includes France, Sweden, Norway, Luxembourg, and others, community 3 includes Senegal, Cyprus, Bahrain, and other, and community 4 includes the United Arab Emirates, Mauritius, and other jurisdictions. Although the United Arab Emirates has a high centrality value (Table 4), the size of the community to which it belongs is small (Table 7). Many European countries belong to Community 1 or 2. This may be because the EU has issued a “Royalty Directive” (2003/49/EC), which exempt corporations from paying withholding tax on royalties paid within the EU member jurisdictions. There are 43 vertices not having edges with other vertices. Regarding these vertices, the same thing can be said as for the dividends (5.2.1 Dividends).

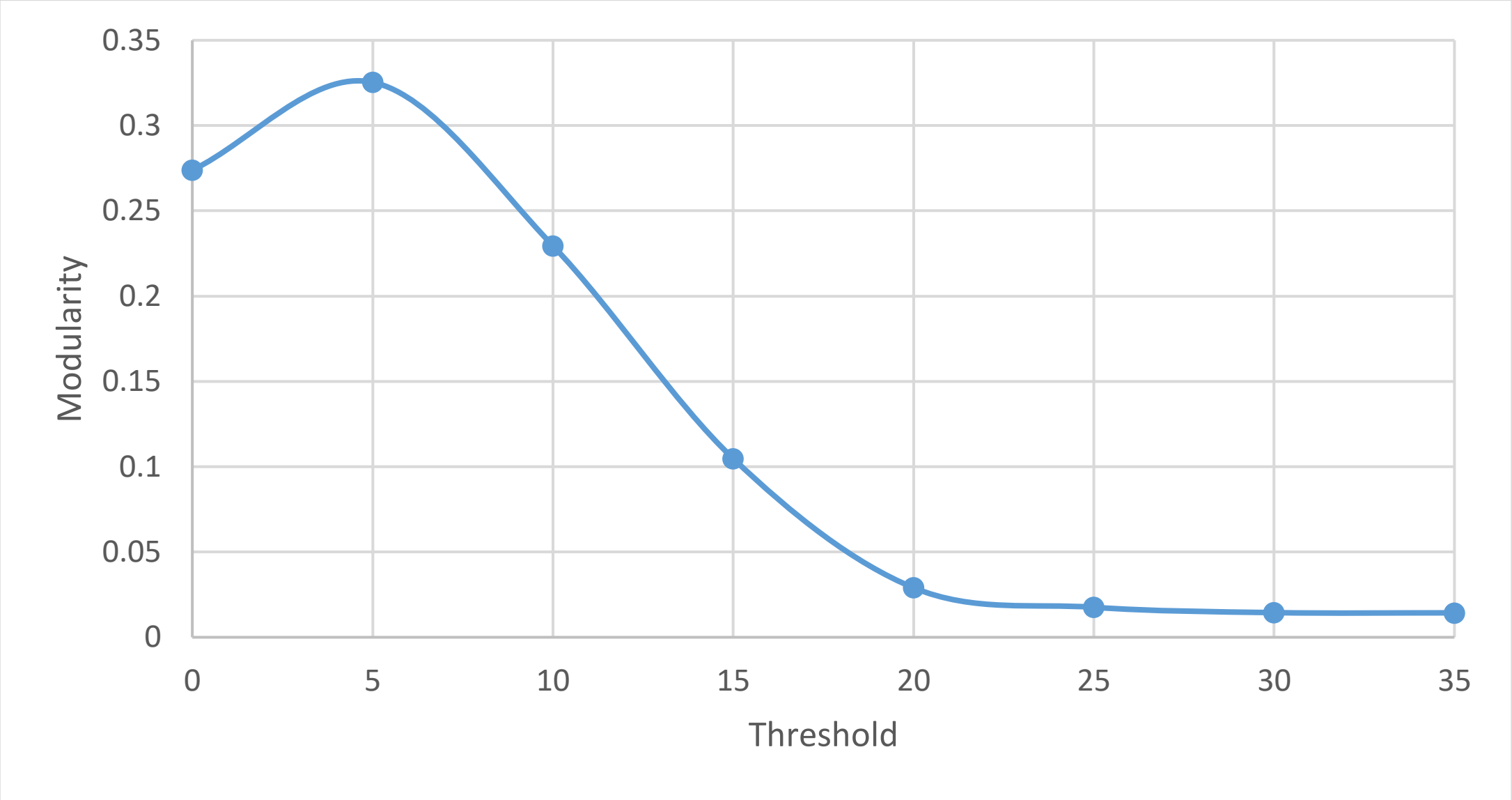

**Fig. 12** Change in modularity, threshold 35 – threshold 0 (in the case of royalties)

**Table 7** Communities (royalties)

| **Community 1** | | **Community2** | | **Community 3** | | **Community 4** | | **no link** | |
|---|---|---|---|---|---|---|---|---|---|
| jurisdiction | centrality | jurisdiction | centrality | jurisdiction | centrality | jurisdiction | centrality | jurisdiction | centrality |
| Switzerland | 0.061055 | France | 0.046474 | Senegal | 0.029019 | UAE | 0.06167 | Afghanistan | 0 |
| Hungary | 0.03975 | Sweden | 0.040768 | Cyprus | 0.028238 | Mauritius | 0.045365 | Angola | 0 |
| Netherlands | 0.030205 | Norway | 0.028039 | Bahrain | 0.015465 | South | 0.00097 | Bolivia | 0 |
| Ireland | 0.028402 | Luxembourg | 0.025179 | Monaco | 0.009766 | Malaysia | 2.93E-05 | Botswana | 0 |
| Malta | 0.019473 | US | 0.019266 | Guernsey | 0.007469 | Oman | 1.87E-05 | Brazil | 0 |
| Slovak | 0.006738 | Latvia | 0.016999 | Macau | 0.006517 | Indonesia | 0 | Brunei | 0 |
| Austria | 0.006683 | UK | 0.012943 | Libya | 0.004796 | Madagascar | 0 | Cambodia | 0 |
| Canada | 0.006667 | Japan | 0.008448 | Liechtenstein | 0.003452 | Mozambique | 0 | Cape | 0 |
| Zambia | 0.006121 | Korea | 0.008326 | Bermuda | 0.003191 | Namibia | 0 | Colombia | 0 |
| Germany | 0.004096 | Italy | 0.007662 | Kuwait | 0.002856 | | | CongoDR | 0 |
| Spain | 0.003763 | Russian | 0.004358 | Nicaragua | 0.001707 | **Community 5** | | Costa | 0 |
| Finland | 0.002521 | Tunisia | 0.001673 | Paraguay | 0.001707 | jurisdiction | centrality | DominicanR | 0 |
| Belgium | 0.001414 | Estonia | 0.001627 | Puerto | 0.001707 | Cameroon | 0.00016 | El | 0 |
| Ukraine | 0.001283 | Denmark | 0.001546 | Martin | 0.001707 | Gabon | 0.024003 | Equatorial | 0 |
| Czech | 0.000641 | Greece | 0.000856 | Suriname | 0.001707 | CongoR | 7.79E-05 | Fiji | 0 |
| Singapore | 0.000521 | Croatia | 0.000776 | Georgia | 0.001528 | Chad | 4.57E-05 | Ghana | 0 |
| Israel | 0.000386 | Bulgaria | 9.30E-05 | China | 0.001124 | | | Guam | 0 |
| Iceland | 0.000135 | Macedonia | 8.72E-05 | Man | 0.000871 | | | Guatemala | 0 |
| Belarus | 0.000118 | Swaziland | 2.58E-05 | Jersey | 0.000871 | | | Guinea | 0 |
| Poland | 7.41E-05 | Saudi | 8.63E-06 | Cote | 0.00054 | | | Guyana | 0 |
| Kyrgyzstan | 6.17E-05 | Ecuador | 6.92E-06 | HK | 0.000318 | | | Honduras | 0 |
| Uzbekistan | 4.15E-05 | Lithuania | 4.90E-06 | Bahamas | 0.000286 | | | India | 0 |
| Bosnia | 3.56E-05 | Panama | 2.51E-06 | Virgin | 0.000286 | | | Iraq | 0 |
| Armenia | 2.49E-05 | Sri | 2.08E-06 | Gibraltar | 0.000286 | | | Jamaica | 0 |
| Pakistan | 2.04E-05 | Algeria | 1.86E-05 | Sao | 0.000286 | | | Kazakhstan | 0 |
| Slovenia | 1.41E-05 | Australia | 0 | Sint | 0.000286 | | | Lebanon | 0 |
| Moldova | 1.41E-05 | Chile | 0 | Cayman | 0.000278 | | | Lesotho | 0 |
| Kosovo | 1.34E-05 | Jordan | 0 | Curacao | 0.000278 | | | Malawi | 0 |
| Egypt | 1.25E-05 | Morocco | 0 | Seychelles | 9.93E-05 | | | Maldives | 0 |
| Mongolia | 9.67E-06 | NZ | 0 | Bonaire | 0 | | | Mexico | 0 |
| Romania | 7.09E-06 | Portugal | 0 | Ethiopia | 0 | | | Myanmar | 0 |
| Montenegro | 5.50E-06 | Trinidad | 0 | Laos | 0 | | | Nigeria | 0 |
| Serbia | 5.50E-06 | Turkey | 0 | Mauritania | 0 | | | Mariana | 2.75E-05 |
| Venezuela | 1.17E-06 | | | Palestinian | 0 | | | Papua | 0 |
| Lucia | 1.13E-06 | | | Qatar | 0 | | | Peru | 0 |
| Albania | 1.04E-06 | | | | | | | Philippines | 0 |
| Argentina | 0 | | | | | | | Rwanda | 0 |
| Azerbaijan | 0 | | | | | | | Ssudan | 0 |
| Taiwan | 0 | | | | | | | Tanzania | 1.39E-06 |
| Uruguay | 0 | | | | | | | Thailand | 0 |
| Vietnam | 0 | | | | | | | Uganda | 0 |
| | | | | | | | | Uvirgin | 0 |
| | | | | | | | | Zimbabwe | 0 |

**6 Discussion**

To resolve the vulnerable elements of tax treaties, it is necessary to amend them. The realm of international tax law is trying to deal with the weak links that arise from a single network that tax laws and tax treaties of each country form, in two ways. The first approach involves multilateral treaties and the second involves peer review.

The multilateral treaty aims to amend tax treaties of each country uniformly, instead of sequentially revising tax treaties through bilateral negotiations as in the past. When a jurisdiction ratifies a multilateral treaty, the new provisions stipulated in the multilateral treaty overwrite the existing provisions of any tax treaties that jurisdiction has concluded, producing the same effect as that of an amendment to the tax treaty. In order for many jurisdictions to ratify a multilateral treaty, the multilateral treaty can effectively include only the minimum standard deemed necessary for the prevention of treaty abuse. From the intention of trying to be “fair” to each jurisdiction, it seems possible only to make revisions to each jurisdiction’s tax treaties “uniformly.” However, according to this study, only some jurisdictions are conducive to treaty shopping, so to solve treaty abuse, it seems sufficient to amend the tax treaties of those jurisdictions for the time being. This would mean fewer jurisdictions that need to obtain consensus, which we believe would allow additional provisions to be introduced into the tax treaties[3].

Peer review means that jurisdictions monitors each other to determine whether provisions are being adequately enforced in each jurisdiction. Again, according to this study, there is a possibility that effective monitoring can be carried out when focusing the monitoring on jurisdictions likely to be used for treaty shopping.

Moreover, international tax avoidance using tax havens is also drawing attention. Because treaty shopping is generally used for shifting companies’ profits to tax havens (Picciotto 1992), resolving treaty

[3] However, seeking prevention measures only for specific countries also has the risk of disrupting efforts towards international tax avoidance. About how the harmful tax competition taking the initiative by the OECD was abandoned, see Sharman (2006).

shopping may also resolve international tax avoidance using tax havens.

## 7 Conclusion

In this study, we were able to clarify which jurisdictions are likely to be used for treaty shopping from the viewpoint of the withholding tax rates, and the relationships between jurisdictions that are likely to be used for treaty shopping and jurisdictions which give rise to a motivation to undertake treaty shopping. However, it cannot be determined from withholding tax rates alone whether these jurisdictions are being used for treaty shopping in reality. As a future task, we would like to study the relationship between withholding tax rates and the economic activities of multinationals by using corporate data.

**Conflict of interest statement** On behalf of all authors, the corresponding author states that there is no conflict of interest.

**Acknowledgement** The present study was supported in part by the Ministry of Education, Science, Sports, and Culture, Grants-in-Aid for Scientific Research (B), Grant No. 17904923 (2017-2019) and (C), Grant No. 26350422 (2014-16). This study was also supported by MEXT as Exploratory Challenges on Post-K computer (Studies of Multi-level Spatiotemporal Simulation of Socioeconomic Phenomena).